\documentclass[5p,authoryear]{elsarticle}

\usepackage{lineno}
\modulolinenumbers[5]
\usepackage[colorlinks]{hyperref}
\usepackage{microtype}
\usepackage{units}
\usepackage{booktabs,multirow}
\usepackage{graphicx}
\usepackage{sidecap}

\usepackage{amsmath}
\usepackage{amssymb}
\usepackage{mathtools}

\renewcommand{\vec}[1]{\mathbf{#1}}
\newcommand{\mat}[1]{\mathrm{#1}}

\usepackage{etoolbox}
\makeatletter
\patchcmd\@combinedblfloats{\box\@outputbox}{\unvbox\@outputbox}{}{%
   \errmessage{\noexpand\@combinedblfloats could not be patched}%
}%
\makeatother
\journal{arXiv}

\begin{document}

\begin{frontmatter}

\title{Scattered slice SHARD reconstruction for motion correction \\in multi-shell diffusion MRI}

\author[cdb,bme,kul]{Daan~Christiaens\corref{corauthor}}
\cortext[corauthor]{Corresponding author}
\ead{daan.christiaens@kcl.ac.uk}

\author[cdb,bme,etsi]{Lucilio~Cordero-Grande}
\author[cdb,bme]{Maximilian~Pietsch}
\author[cdb,bme]{Jana~Hutter}
\author[cdb,bme]{Anthony~N.~Price}
\author[cdb]{Emer~J.~Hughes}
\author[cdb]{Katy Vecchiato}
\author[cdb,bme]{Maria~Deprez}
\author[cdb,mrc]{A.~David~Edwards}
\author[cdb,bme]{Joseph~V.~Hajnal}
\author[cdb,bme]{J-Donald~Tournier}

\address[cdb]{Centre for the Developing Brain, School of Biomedical Engineering \& Imaging Sciences, King's College London, London, U.K.}
\address[bme]{Biomedical Engineering Department, School of Biomedical Engineering \& Imaging Sciences, King's College London, London, U.K.}
\address[kul]{Department of Electrical Engineering, ESAT/PSI, KU Leuven, Leuven, Belgium}
\address[etsi]{Biomedical Image Technologies, ETSI Telecomunicación, Universidad Politécnica de Madrid \& CIBER-BBN, Madrid, Spain}
\address[mrc]{MRC Centre for Neurodevelopmental Disorders, King's College London, London, U.K.}

\tnotetext[abbrev]{List of Abbreviations: 
dHCP, developing Human Connectome Project;
dMRI, diffusion magnetic resonance imaging; 
EPI, echo planar imaging;
FWHM, full width half maximum;
GMM, Gaussian mixture model;
ODF, orientation distribution function; 
RMSE, root-mean-square error; 
SH, spherical harmonics; 
SHARD, spherical harmonics and radial decomposition;
SVD, singular value decomposition;
SNR, signal-to-noise ratio;
SSP, slice sensitivity profile.
}

\begin{abstract}
Diffusion MRI offers a unique probe into neural microstructure and connectivity in the developing brain. However, analysis of neonatal brain imaging data is complicated by inevitable subject motion, leading to a series of scattered slices that need to be aligned within and across diffusion-weighted contrasts. 
Here, we develop a reconstruction method for scattered slice multi-shell high angular resolution diffusion imaging (HARDI) data, jointly estimating an uncorrupted data representation and motion parameters at the slice or multiband excitation level. The reconstruction relies on data-driven representation of multi-shell HARDI data using a bespoke spherical harmonics and radial decomposition (SHARD), which avoids imposing model assumptions, thus facilitating to compare various microstructure imaging methods in the reconstructed output. Furthermore, the proposed framework integrates slice-level outlier rejection, distortion correction, and slice profile correction. 
We evaluate the method in the neonatal cohort of the developing Human Connectome Project (650 scans). Validation experiments demonstrate accurate slice-level motion correction across the age range and across the range of motion in the population. Results in the neonatal data show successful reconstruction even in severely motion-corrupted subjects. In addition, we illustrate how local tissue modelling can extract advanced microstructure features such as orientation distribution functions from the motion-corrected reconstructions.
\end{abstract}


\end{frontmatter}

%\linenumbers

%%%%%%%%%%%%%%%%%%%%%%%%%%%%%%%%%%%%%%%%%%%%%%%%%%%%%%%
%%%%%%%%%%%%%%%%%%%%%%%%%%%%%%%%%%%%%%%%%%%%%%%%%%%%%%%

\section{Introduction}

Diffusion magnetic resonance imaging (dMRI) offers a unique probe into brain microstructure and connectivity throughout the lifespan \citep{LeBihan1986}. However, the relatively long acquisition time of modern dMRI protocols leads to inevitable subject motion, particularly in pediatric and geriatric cohorts. Motion correction is therefore a crucial processing step in dMRI analysis.

The first approaches to dMRI motion correction operated on a volume level, retrospectively seeking a rigid transformation for each image volume that describes the head motion of the subject during the scan \citep{Rohde2004, Andersson2016}. In this case, volume-level motion correction can be regarded as a multi-contrast image registration problem, in which the acquired image volumes are interpolated into the moving subject reference frame. However, the vast majority of dMRI data are collected as stacks of slices using echo planar imaging (EPI) sequences \citep{Turner1990}. Subject motion can hence occur between slices, as much as between volumes, leading to substantial intra-volume motion artefacts in less compliant subject groups. In such groups, such as the neonatal cohort used in this work, effective motion correction needs to realign the individual slices into a self-consistent image. 

The slice is the natural unit of data for motion correction in EPI. Indeed, subject motion can be assumed frozen within the short readout time of every EPI shot (a single slice or a group of slices acquired simultaneously in a single multiband excitation). Inter-shot motion scatters the slice positions with respect to the subject, which can lead to undersampled areas. Intra-shot motion induces a linear phase shift \citep{Anderson1994}, which can give rise to slice dropouts that need to be treated as outliers in further analysis. Crucially, both sources of missing data impede direct interpolation of the acquired image data in the subject reference. 

Furthermore, the diffusion-weighted contrast depends on the acquired $b$-value and gradient orientation in the scanner reference frame. If the subject moves during the scan, the orientation-dependent dMRI contrast needs to be modulated to match the encoding direction in the subject reference frame. In volume-level dMRI motion correction, this reduces to a straightforward reorientation of the diffusion gradient encoding vectors \citep{Leemans2009}. However, in slice-level motion correction, individual slices within any given volume may be differently affected by motion-induced rotation, introducing scatter in the effective diffusion gradient direction \citep{Oubel2012, Fogtmann2014}. In other words, each shot may have seen a unique diffusion encoding gradient. As such, it is no longer sufficient to reorient the diffusion encoding of the volume, nor is it correct to interpolate across slices that have a different diffusion sensitization. 

Therefore, slice-level motion correction requires a conceptual shift to an inverse problem perspective, in which the output image and slice motion parameters are jointly optimized to best explain the acquired scattered slice data. In anatomical imaging, where such approach is known as \emph{slice-to-volume reconstruction} \citep{Rousseau2006, Gholipour2010, KuklisovaMurgasova2012}, a single (often super-resolved) image volume is fitted to multiple orthogonally-acquired slice stacks. In diffusion imaging, where data is collected with varying image contrast, a multi-dimensional signal representation is fitted to the set of slices, both within and across the various diffusion encodings. As such, the issues with interpolation are circumvented in favour of a $q$-space data representation that can be evaluated for any $b$-value and reoriented gradient direction.

To facilitate a wide range of further analyses, the signal representation must be model-free, whilst also offering rank-reduction properties needed to make the data self-consistent. Prior work in dMRI slice-to-volume reconstruction has used representations based on the diffusion tensor \citep{Jiang2009, Fogtmann2014, Marami2016, Marami2017}, multi-tensor models \citep{Marami2018}, single-shell spherical harmonics \citep{Deprez2019}, or Gaussian processes \citep{Scherrer2012, Andersson2017}, which can limit the supported input data or downstream analysis. Here, we use a data-driven signal representation for multi-shell dMRI based on spherical harmonics and a radial decomposition (SHARD) that was shown to have compelling low-rank properties highly suitable for motion correction applications \citep{Christiaens2018}. Its model-free nature enables fair comparison between various microstructure models and data analysis methods.

The proposed slice-to-volume reconstruction framework corrects subject motion and static EPI distortion, and also incorporates the slice profile to ensure through-plane resolution recovery. Furthermore, we integrate intensity-based slice outlier detection based on a log-normal mixture model. We validate accuracy in brain dMRI data with simulated motion traces and measure test-retest reproducibility in scans with and without motion. In addition, we apply and evaluate this method in a cohort of 650 neonatal scans acquired as part of the developing Human Connectome Project (dHCP). Our software implementation is publicly available as a module for MRtrix3 \citep{MRtrix}. The processed neonatal data will be part of the next dHCP release.

%%%%%%%%%%%%%%%%%%%%%%%%%%%%%%%%%%%%%%%%%%%%%%%%%%%%%%%

\section{Method}

\subsection{Signal representation and inverse problem}

\subsubsection{SHARD representation}

Given multi-shell HARDI data acquired in a set of slices or multiband excitations, we wish to estimate the unknown motion traces and reconstruct the motion-corrected signal. Due to the orientation-dependence of the dMRI contrast, this reconstruction requires a model-independent signal representation that spans across the angular as well as radial $q$-space domain. To this end, we have recently introduced a data-driven representation named Spherical Harmonics And Radial Decomposition (SHARD) \citep{Christiaens2018}, which links the conventional spherical harmonics (SH) representation in each shell with a bespoke singular value decomposition (SVD) in the radial domain to represent multi-shell dMRI data as a linear combination of orthogonal components. This decomposition was shown to outperform alternative signal representations whilst also capturing all micro- and meso-structural information in the signal up to a chosen basis rank $r$. Furthermore, cross-validation experiments also showed that SHARD facilitates efficient rank reduction, which can help ensure robust volume-to-slice registration.

The target image is thus represented as a 4-D image of SHARD reconstruction coefficients, encoded in vector $\vec{x}$ of length $n_{xyz} \cdot r$, where $n_{xyz} = n_x \cdot n_y \cdot n_z$ is the image grid size of the reconstruction. Given this representation and subject motion parameters, a linear forward model enables predicting the expected dMRI contrast in each slice. The inverse problem then consists of optimizing the target $\vec{x}$ for maximum similarity to the acquired scattered slice data, whilst simultaneously finding the subject motion traces that encode the slice positions.

\subsubsection{Forward model}\label{sec:forward}

\begin{figure}[t]
	\centering
	\includegraphics[width=\columnwidth]{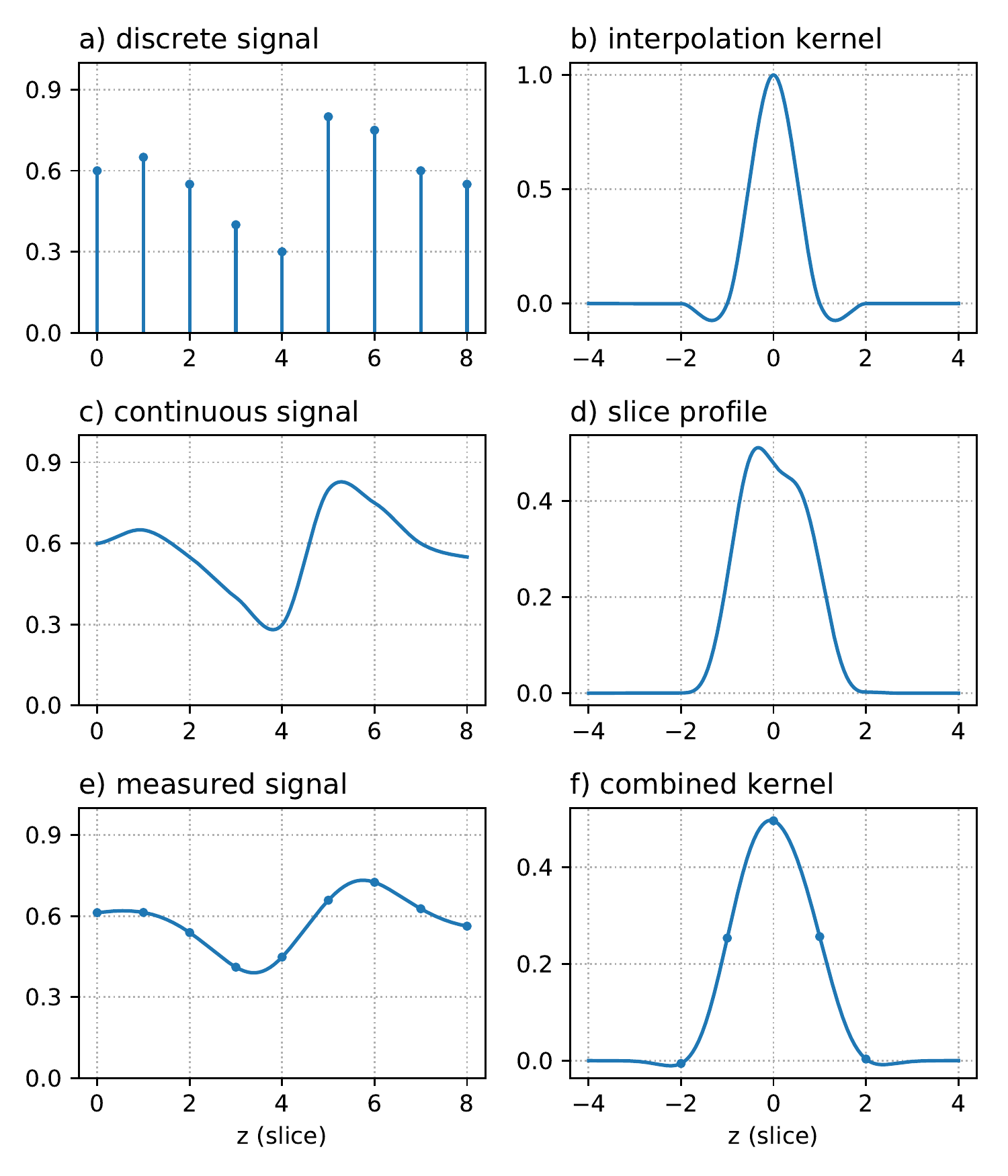}
	\caption{Slice profile operator in the forward model: a) example reconstructed signal on a discrete grid; b) cubic interpolation kernel; c) continuous signal represented by $a$, modelled as the convolution of $a$ and $b$; d) actual slice sensitivity profile in the neonatal dHCP data, i.e., the (normalized) transverse magnetization $M_{xy}(z)$ calculated using a Bloch equation simulator taking into account the properties of the multiband RF excitation, spatial interleaving pattern of the slices, and sequence timings; e) prediction of the measured signal as the convolution of $c$ and $d$; f) combined convolution kernel used in the reconstruction, namely the convolution of $b$ and $d$ sampled at the slice separation distance. The blurring and slice selection matrix $\mat{B}_s$ implements a discrete convolution with this combined kernel $f$, which is mathematically equivalent but computationally more efficient.}\label{fig:ssp}
\end{figure}

For given rigid motion parameters $\mu_s$ of the subject at the time point of acquiring slice $s$, a prediction $\hat{\vec{y}}_s$ of the acquired signal in that slice can be obtained as
\begin{equation}
	\hat{\vec{y}}_s = \mat{B}_s \, \mat{M}(\mu_s) \, \mat{Q}_s(\mu_s) \;\vec{x}        \quad ,
\end{equation}
where
\begin{itemize}
\item $\mat{Q}_s(\mu_s)$ encodes the SHARD basis for $q$-space, used to project the reconstruction coefficients $\vec{x}$ to the dMRI contrast for the reoriented diffusion encoding of slice $s$. This is a block diagonal matrix of size $n_{xyz} \times n_{xyz} \cdot r$ in which the multi-shell signal basis of size $1 \times r$ is repeated on the diagonal. The signal basis depends on the diffusion encoding of slice $s$ ($b$-value and direction) and also on the motion parameters $\mu_s$ due to the need for gradient reorientation.
\item $\mat{M}(\mu_s)$ is a linear motion operator, mapping the projected dMRI contrast to scanner space. This is a square matrix of size $n_{xyz} \times n_{xyz}$, in which every row interpolates scanner grid positions in subject space for the current motion $\mu_s$. This work uses cubic interpolation, leading to a sparse convolution structure \citep{Keys1981}. $\mat{M}(\mu_s)$ can also account for EPI distortion (cf.\,Section~\ref{sec:distortion}).
\item $\mat{B}_s$ is a blurring and slice selection matrix associated with the slice sensitivity profile (SSP). This operation convolves the simulated contrast with the SSP in the scanner slice direction $z$, and evaluates the result at the location of slice $s$, resulting in a sparse matrix of size $n_{xy} \times n_{xyz}$. SSP convolution is explained in more detail in Fig.~\ref{fig:ssp}. If the exact SSP is not known, it can be approximated with a Gaussian point spread function of full width half maximum (FWHM) matched to the slice thickness.
\end{itemize}
This forward model enables generating predictions for the signal in a scattered collection of slices, each with their own motion state and dMRI encoding. The slice profile matrix also enables super-resolution of overlapping slices in the through-plane direction.

To shorten the notation below, we define the full reconstruction matrix for every slice $s$ as
\begin{equation}
	\mat{R}_s(\mu) = \mat{B}_s \, \mat{M}(\mu_s) \, \mat{Q}_s(\mu_s) 	\quad,
\end{equation}
with $\mu$ denoting the total subject motion trace. Note that $\mu_s$ captures the position of the \emph{subject} at the time of acquiring slice $s$; individual slices acquired simultaneously in a single multiband excitation thus share the same motion state.

\subsubsection{Inverse problem}

Given the acquired scattered slice data, we seek the optimal reconstruction $\vec{x}^\ast$ and rigid motion parameters $\vec{\mu}^\ast$ that maximize similarity between the acquired data and its prediction. Assuming Gaussian noise, the cost function is the weighted $L_2$-norm of the residuals in all slices $\vec{y}_s$:
\begin{equation}\label{eq:inverseproblem}
	\min_{\vec{x}, \vec{\mu}} \sum_s \tfrac{w_s}{n_v} \, \| \vec{y}_s - \mat{R}_s(\mu) \, \vec{x} \|^2 + \pi(\vec{x})   \quad ,
\end{equation}
where $w_s$ is a slice weight that is used for outlier rejection, $n_v$ is the number of acquired volumes, and $\pi(\vec{x})$ is a regularization term on the reconstruction. Regularization is needed to stabilize noise amplification in the SSP deconvolution and inverse interpolation. In this work we define the regularizer as
\begin{equation}
	\pi(\vec{x}) = \lambda^2\,\| \mat{L}\,\vec{x} \|^2 + \zeta^2\,\| \mat{Z}\,\vec{x} \|^2 	\quad,
\end{equation}
where the first term is an isotropic Laplacian convolution filter and the second term is an 8th-order derivative filter in the slice direction. The derivative filter's high-pass characteristic helps stabilize the ill-conditioned slice profile deconvolution at reasonable computational cost.

The optimization process alternates between a \emph{reconstruction} step in which the SHARD representation $\vec{x}$ is optimized given a current motion estimate $\vec{\mu}$, and a \emph{registration} step in which the motion parameters are updated for the current $\vec{x}$. In addition, an \emph{outlier reweighting} step is incorporated to discard damaged slices from the reconstruction. These steps are described in detail in sections \ref{sec:reconstruction}--\ref{sec:orweighting}, and their integration is described in section \ref{sec:integration} and illustrated in Fig.~\ref{fig:recscheme}.

\subsubsection{Modelling EPI distortion}\label{sec:distortion}

EPI distortion due to magnetic susceptibility and eddy currents introduce an additional geometric transformation in the data as a displacement field along the phase encoding direction in every slice. These can, to a good first order approximation, be described by an invariant B0 field map in the \emph{subject} reference frame \citep{Andersson2003}. Under subject motion, the field map is thus assumed to move with the subject's head. Given current motion parameters, the rigidly-aligned field describes the pull-back deformation needed to map distorted dMRI slices into undistorted, but still motion-corrupted space. A forward model of motion and distortion hence needs to invert the field map upon each update of the motion parameters.

The effect of EPI distortion should in principle be modelled as part of the geometric transformation in matrices $\mat{M}(\mu_s)$. Although our implementation supports this, we found that the benefits of this direct model do not outweigh the significant computational cost of the required field inversion in every iteration of the reconstruction step. Instead, the inverse problem solver unwarps the input dMRI data before each reconstruction step, given the field map and the current estimate of the subject motion trace and accounting for the appropriate Jacobian scaling \citep{Andersson2003}. As such, the implementation achieves higher run time efficiency at the expense of a double interpolation in the reconstruction process.

\subsection{Reconstruction}\label{sec:reconstruction}

During \emph{reconstruction} steps, we solve the sub-problem:
\begin{multline}\label{eq:reconstruction}
	\vec{x}^{(k)} = \arg\min_{\vec{x}} \tfrac{1}{n_v} \| \mat{W} (\vec{y} - \mat{R}^{(k)} \,\vec{x}) \|^2 \\+ \lambda^2\,\| \mat{L}\,\vec{x} \|^2 + \zeta^2\,\|\mat{Z}\,\vec{x}\|^2   \quad,
\end{multline}
where $\mat{R}^{(k)} = [ \cdots | R_s^{\top(k)} | \cdots ]^\top$ is the combined reconstruction matrix of all scattered slices in the current iteration $k$, and $\mat{W} = \text{diag}(\cdots \sqrt{w_s} \cdots)$. This is a large, sparse least-squares fit to all combined scattered slice data.

\paragraph{Least-squares system} The reconstruction problem \eqref{eq:reconstruction} can be written as the least-squares solution to 
\begin{equation}
	\left[ \begin{array}{c}
		\vdots \\ \sqrt{w_s}\,\vec{y}_s \\ \vdots \\ \hline 0 \\ \hline 0
	\end{array} \right] = \left[ \begin{array}{c}
		\vdots \\ \sqrt{w_s}\,\mat{R}_s \\ \vdots \\ \hline \lambda\,\mat{L} \\ \hline \zeta\,\mat{Z} 
	\end{array} \right] \; \vec{x} \quad,
\end{equation}
where the index $(k)$ was dropped from the notation. The normal equations of this system are
\begin{equation}\label{eq:normeq}
	\sum_s w_s\,\mat{R}_s^\top\vec{y}_s = \left( \sum_s w_s\,\mat{R}_s^\top\mat{R}_s + \lambda^2\,\mat{L}^\top\mat{L} + \zeta^2\,\mat{Z}^\top\mat{Z} \right) \, \vec{x} \quad.
\end{equation}
The system matrix dimension is in the order of 100 million, too large to fit in memory and invert on normal hardware. Therefore, we rely on an iterative conjugate gradient (CG) method that takes advantage of the sparse structure of the matrices involved.

\paragraph{CG solver} Conjugate gradient is an iterative numerical method to solve a square symmetric linear system of equations $\vec{z} = \mat{E}\,\vec{x}$ such as the one in \eqref{eq:normeq}. Rather than requiring the inverse or Cholesky decomposition of matrix $\mat{E}$, the CG solver only requires access to the matrix multiplication $\mat{E}\,\vec{r}$, which can be implemented without ever needing to construct $\mat{E}$ explicitly. Instead, our implementation chains up the linear operators $\mat{Q}_s$, $\mat{M}$, $\mat{B}_s$, and their transpose, as well as the regularization terms $\mat{L}$ and $\mat{Z}$, as a series of image filters and interpolation steps. As such, we avoid needing to store and invert the full reconstruction matrix, saving otherwise prohibitive memory load by leveraging the sparse structure in a custom matrix-free solver. Furthermore, the iterative nature of the CG solver allows us to relax the convergence criteria in early epochs, thus saving on run time when the motion parameters have not yet converged.

\subsection{Registration}\label{sec:registration}

During \emph{registration} steps, the rigid motion parameters are updated by optimizing spatial alignment between a rank-reduced reconstruction $\tilde{\vec{x}}^{(k)}$ and the acquired scattered slice data. To this end, we apply rigid image registration between the predicted dMRI contrast and the acquired slices to update the motion parameters at each time point by solving the associated sub-problem
\begin{equation}\label{eq:registration}
	\mu_s^{(k+1)} = \arg\min_{\mu_s,\alpha_s} \| \vec{y}_s - \alpha_s \mat{R}_s(\mu_s) \, \tilde{\vec{x}}^{(k)} \|^2   \quad,
\end{equation}
where $\alpha_s$ is an intensity scale factor (details below). Subsequently, we filter the rigid motion parameters for outlier robustness and temporal consistency. In all processing, the rigid motion parameters are represented in Lie algebra, as described below.

\paragraph{Rigid motion parameterization}
The rigid pose of each slice in material coordinates is characterized by 6 degrees of freedom: 3 translation parameters and 3 rotation parameters. Conventional parameterizations of rigid motion in Euler angles or quaternions are either degenerate (gimbal lock) or overparameterized ($>6$ parameters), complicating the calculation of the Jacobian in the image registration process. We therefore chose to parameterize the rigid motion group SE(3) by its corresponding Lie algebra se(3), thus linearizing the manifold \citep{Blanco2010}. In this representation, the pose vector $\mu_s = [ t_x, t_y, t_z, r_x, r_y, r_z ]^\top$ uniquely characterizes a $4 \times 4$ rigid transformation matrix $\mathcal{T}(\mu_s)$ in homogeneous scanner coordinates under the following mapping:
\begin{equation}\label{eq:liemap}
	\mathcal{T}(\mu_s) = \exp \begin{pmatrix} 0 & -r_z & r_y & t_x \\ r_z & 0 & -r_x & t_y \\ -r_y & r_x & 0 & t_z \\ 0 & 0 & 0 & 0  \end{pmatrix} \quad,
\end{equation}
where $\exp$ is the matrix exponential. The inverse mapping can be achieved with the matrix logarithm. This linearized representation is well suited for direct optimization on the SE(3) manifold, such as rigid registration \citep{Blanco2010}.

\paragraph{Image registration}
The nonlinear least-squares problem in \eqref{eq:registration} is a volume-to-slice(s) rigid registration problem under a sum-of-squared-differences similarity metric, in which the rigid pose $\mu_s$ of the subject at the acquisition of slice $s$ is optimized for best alignment with the predicted \emph{rank-reduced} dMRI contrast in the \emph{reoriented} diffusion encoding. The auxiliary unknown $\alpha_s$ enables scaling the predicted image intensity to slices with partial saturation artefacts, hence improving robust alignment in the presence of intensity dropouts and spin history effects. Note that these scale factors are \emph{not} used in the reconstruction (equation~\eqref{eq:reconstruction}); ultimately, slices with intensity dropout are treated as outliers in the forward model.

Our implementation first generates a 3-D prediction of the reoriented dMRI contrast based on the motion parameters $\mu_s^{(k)}$ of the previous iteration, thus approximating the reoriented contrast in the interest of computational efficiency. The subsequent rigid volume-to-slice alignment is then solved with the Levenberg-Marquardt algorithm \citep{Marquardt1963}. The Jacobian $\frac{\partial}{\partial\mu_s} R_s(\mu_s)\,\vec{x}^{(k)}$ is expressed analytically, based on the exact Jacobian of the composition of a point and a pose in Lie algebra as outlined in \citet{Blanco2010}.

\paragraph{Temporal filter}
The independent, per-slice registration described above is only as good as its target and source images. The target slice can be affected by signal dropout where there is no meaningful contrast to register to. The source image is the intermediate result of a dynamic motion correction process and can therefore likewise be affected by yet unresolved slice dropouts and misalignment. To mitigate these effects, we apply minimal temporal filtering to the estimated motion traces. To this end, we index the motion parameters $\mu_t$ according to their acquisition time point $t$, which depends on the volume order and on the slice interleave and shift pattern (e.g.\,odd-even slice order). The filter operates in two stages. The first stage aims to replace untrustworthy motion parameters in outlier slices with the mean pose of their neighbours. This is implemented with the weighted average
\begin{equation}
	\mu'_t = w_t \, \mu_t + (1-w_t) \, \tfrac{1}{2} (\mu'_{t-1} + \mu'_{t+1}) \quad,
\end{equation}
where $w_t = w_s(t)$ are the outlier weights of the corresponding slices as estimated from the image similarity between data and prediction indexed in temporal order. This filter is calculated by solving a square band diagonal linear system for the vector of unknowns $[ \, \cdots \, \mu'_{t-1} \; \mu'_{t} \; \mu'_{t+1} \, \cdots \, ]^\top$. The second stage aims for edge-preserving smoothing in the temporal domain to suppress spot noise, implemented as a straightforward median filter with window size fixed to 5 in all experiments.

\subsection{Outlier reweighting}\label{sec:orweighting}

Outlier detection is needed to downweight the influence of dropout or otherwise artefacted slices in the reconstruction. Here, we use a probabilistic criterion based on the intensity difference between the acquired data and its prediction in each multiband excitation, i.e., based on the residuals of the reconstruction step. When the residuals in a slice are large compared to all other slices at similar $b$-value, that slice is detected and discarded as an outlier.

\begin{figure}
	\centering
	\includegraphics[width=\columnwidth]{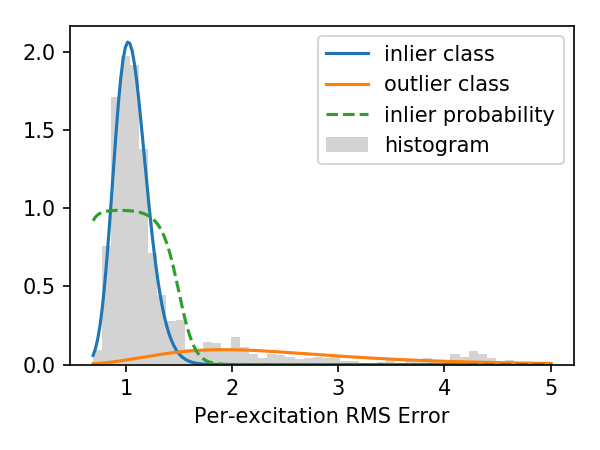}
	\caption{Illustration of the slice outlier reweighting procedure, in this case shown for a representative subject in the $b=\unitfrac[2600]{s}{mm^2}$ shell. A histogram for the root-mean-squared error (RMSE) per multiband excitation is shown in grey. The blue and orange lines plot the log-normal inlier and outlier components respectively. The green dashed line plots the associated inlier probability, showing a sharp separation between inlier and outlier shots.}\label{fig:orgmm}
\end{figure}

The procedure therefore starts by calculating the root-mean-squared error (RMSE) within the brain mask for each multiband excitation. The reliability of each slice is subsequently calculated from a probabilistic mixture model, enabling unsupervised separation in inlier and outlier classes. Since the SNR can be different across shells, this mixture model is fitted to the RMSE of all slices with the same $b$-value. The RMSE in the inlier class is expected to be $\chi$-distributed, albeit with non-constant degrees of freedom due to the variable number of voxels included in the mask. The outlier class is expected to follow a non-central $\chi$-distribution, again with non-constant degrees of freedom. Both distributions can be well approximated with a log-normal distribution, thus simplifying the outlier classification to a two-class Gaussian Mixture Model (GMM) of the log-RMSE. We fit the GMM using Expectation-Maximization \citep{Dempster1977}, with the inlier class mean and standard deviation respectively initialized at the median and median absolute deviation (MAD) log-RMSE, and the outlier class analogously initialized at 3$\times$RMSE. After convergence, the inlier class membership probability determines the weight $w_s$ of each slice. The log-GMM fit is illustrated in Fig.~\ref{fig:orgmm}.

\subsection{Integration}\label{sec:integration}

\begin{figure}
	\centering
	\includegraphics[width=\columnwidth]{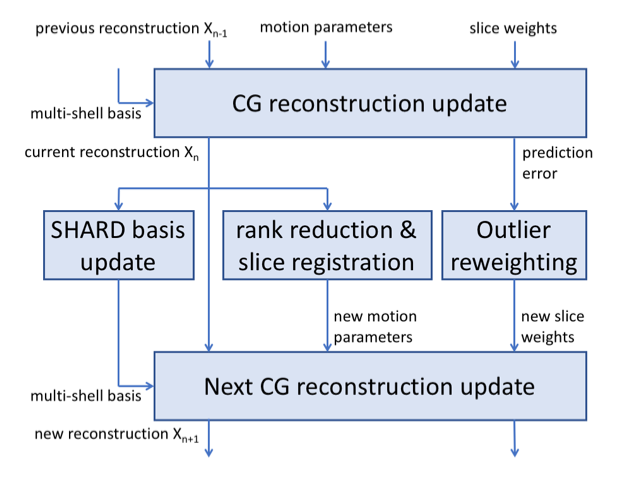}
	\caption{Process integration scheme, representing one epoch in the motion correction framework. Between each major update of the reconstruction, the multi-shell SHARD basis, motion parameters and slice outlier weights are updated and passed to the next iteration.}\label{fig:recscheme}
\end{figure}

The reconstruction, registration, and outlier reweighting steps described in the preceding sections are alternatingly repeated upon every epoch in the combined motion correction framework. Furthermore, the data-driven SHARD basis is also updated at each iteration because the low-rank representation is learned from the data and thus improves as the data becomes more self-consistent. The integration of these steps is schematically shown for a single epoch in Fig.~\ref{fig:recscheme}. Given the current motion parameters and slice weights, and initialized with the previous reconstruction, the CG solver updates the reconstruction and also outputs the residuals between the data and prediction. This current reconstruction is then used to update the SHARD basis and motion parameters, and also to initialize the next reconstruction step. To update the motion parameters, a rank-reduced reconstruction is registered to the slices of the original input data as described above. The outlier weights are updated based on the residuals. On the next epoch, this process repeats with the new inputs.

The iterative motion correction process is initialized with all motion parameters set to 0 (no subject motion) and all slice weights set equal. In the default setup, which we found to be robust across various datasets, the first 2 epochs are constrained to volume-level registration. Afterwards, we proceed with slice-to-volume registration on the multiband excitation level, typically for another 3 epochs. During these stages the reconstruction is terminated after 3 iterations of the CG solver, as the reconstruction will be further refined in later epochs. The last reconstruction step, with final motion parameters, is run for 10 CG iterations to further enhance the precision of the output. Registration is always constrained to 10 iterations of the Levenberg-Marquardt optimizer. To ensure robust alignment, the registration input image (the current rank-reduced reconstruction) is filtered with isotropic spatial smoothing. The FWHM of this smoothing kernel is gradually reduced from 3 to 1 times the voxel size, thus adopting a multi-scale strategy in the motion correction process.

The motion correction method is implemented in the open source software toolset MRtrix3 \citep{MRtrix}. The reconstruction, registration and outlier reweighting processes are implemented as standalone compiled binaries. The integrated motion correction pipeline is implemented as a customizable wrapper script. Our source code is publicly available as an external module (\url{https://gitlab.com/ChD/shard-recon}) that we plan to integrate in the MRtrix3 core.

%%%%%%%%%%%%%%%%%%%%%%%%%%%%%%%%%%%%%%%%%%%%%%%%%%%%%%%

\begin{figure*}
	\centering
	\includegraphics[width=.8\textwidth]{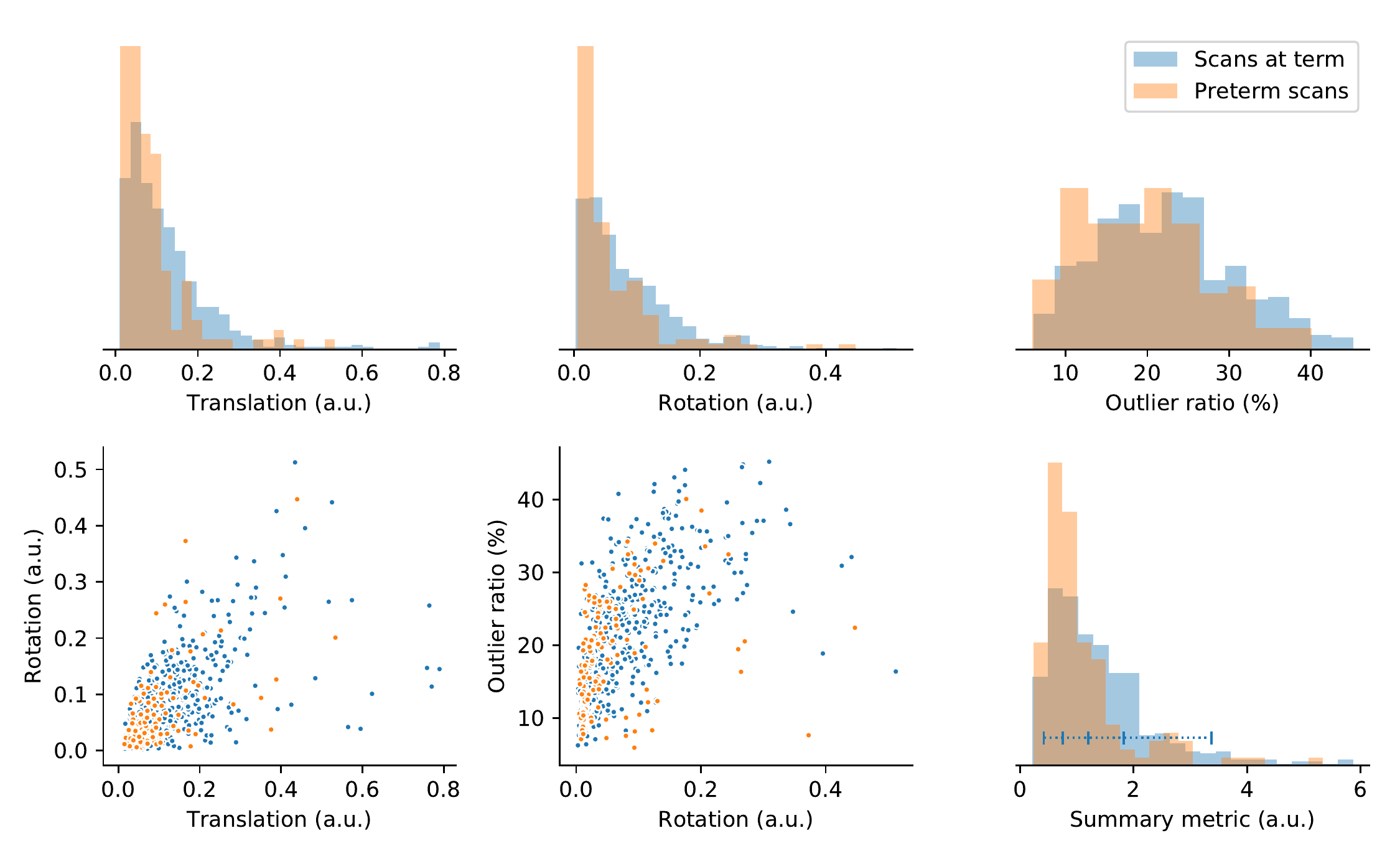}
	\caption{Distribution of motion metrics in the preterm and term cohorts. The top row displays normalized histograms of the calculated translation, rotation and slice outlier metrics, respectively. The bottom left and middle scatter plots show the correlation between these metrics. The bottom right graph displays the histogram of the summary motion metric, calculated from the others. We generally detect less motion in the preterm scans. The blue horizontal bar indicates the percentiles within the term group used to select subjects for display in this article.}\label{fig:statsplot}
\end{figure*}

\section{Data and Experiments}

\subsection{Data and preprocessing}

Neonatal diffusion-weighted brain imaging data were acquired as part of the Developing Human Connectome Project\footnote{\url{http://www.developingconnectome.org/project}} (dHCP), following written parental informed consent. Participants were scanned during natural sleep, in the first week after birth and repeated at term-equivalent age for those born preterm. All study procedures were reviewed and approved by the Riverside Research Ethics Committee (14/LO/1169). For this work, we included all subjects scanned before the end of 2018 that have a complete dMRI acquisition and undamaged $k$-space reconstruction. Partial and interrupted scans were excluded, but subjects with incidental findings were not. The resulting selection comprises 650 scans in 571 unique subjects: 134 preterm scans (post-menstrual age (PMA) at scan between 26 and 37 weeks) and 516 scans at term (between 37 and 44 weeks PMA).

All scans were conducted on a \unit[3]{T} Philips Achieva system using a bespoke 32-channel neonatal head coil and patient-handling system \citep{Hughes2017}. The dMRI acquisition parameters are: TE = \unit[90]{ms}; TR = \unit[3800]{ms}; multiband factor MB=4; SENSE factor~1.2; Partial Fourier factor 0.855; \unit[1.5]{mm} in-plane resolution; \unit[3]{mm} slice thickness with \unit[1.5]{mm} slice overlap \citep{Hutter2017}. The diffusion gradient scheme is sampled across 4 shells with $b = 0,\,400,\,1000,\,\text{and}\,\unitfrac[2600]{s}{mm^2}$, with 20, 64, 88, and 128 samples respectively \citep{Tournier2015}. The slice order uses an interleave factor 3 with shift 2. Volumes are fully interleaved across $b$-values to optimize gradient duty cycle effects, and are furthermore interleaved across 4 orthogonal phase encoding directions to aid in susceptibility distortion correction \citep{Hutter2017, Bastiani2019}.

In addition, a test-retest scan of one adult volunteer was conducted on the same system but using a 32-channel Philips head coil. All acquisition parameters were matched to the neonatal protocol, except the in plane resolution and slice thickness were scaled isotropically to $\unit[2.5 \times 2.5 \times 5.0]{mm}$ (with \unit[2.5]{mm} slice overlap) to accommodate a larger field of view. The volunteer was instructed to stay still during the first scan and to move actively during the rescan.

Prior to motion correction, all data were preprocessed with image denoising \citep{Veraart2016}, Gibbs ringing suppression \citep{Kellner2016}, and B0 field map estimation \citep{Andersson2003}. Image denoising is performed in the complex (phase-magnitude) domain after linear phase removal, where noise can be assumed additive Gaussian. The noise suppression is achieved using SVD truncation based on Marchenko-Pastur theory in $7\times 7\times 7 $ patches \citep{Veraart2016}. The static B0 field map is estimated from the $b=0$ volumes using FSL Topup \citep{Andersson2003}, to enable correcting magnetic susceptibility-induced distortion. This field map, alongside the preprocessed dMRI data and a brain mask \citep{Smith2002}, is input to our software.

\subsection{Setup}

For all experiments in this paper, these scattered neonatal dMRI data are reconstructed in a SHARD basis of rank 89, corresponding to SH order $\ell_\text{max} = 0, 4, 6, 8$ in shells $b = 0, 400, 1000, 2600 \,\unitfrac{s}{mm^2}$ respectively. The registration steps operate at a reduced basis rank 22, i.e., a truncated radial decomposition into respectively 3, 2 and 1 SHARD components in the $\ell=0$, $\ell=2$ and $\ell=4$ SH bands ($3\cdot1+2\cdot5+1\cdot9=22$ basis coefficients). At this rank, we found registration to be sufficiently accurate and also sufficiently robust in the dHCP cohort. The regularization parameters in the reconstruction were empirically tuned to $\lambda = \zeta = 10^{-3}$. We use the slice profile calculated for the specific acquisition sequence as illustrated in Fig.~\ref{fig:ssp}. We also use the multiband factor 4 and the slice interleave order for constructing the temporally ordered motion states.

In addition, we introduce (multiplicative) prior slice weights to flag top and bottom slices as outliers due to asymmetric spin history. These prior inlier probabilities are set to 0 for the top slice, and 0.1 and 0.5 for the two bottom slices, which we experimentally found to be adequate. We also apply a slice-wise intensity scale factor on the preprocessed input data to mitigate a venetian blind artefact mainly affecting the low $b$-value shells. These scale factors are fixed per shell for the entire population, and derived heuristically using the residuals of a low-order polynomial fit in population-averaged raw data.

The proposed slice-to-volume reconstruction took about \unit[90]{min} run time per subject on an 8-core i7 workstation. For a fixed number of iterations, the run time scales linearly with the size of the data and with the reconstructed SHARD basis rank.

\subsection{Motion metrics}

The amount of motion detected in each subject is quantified using summary metrics of the total subject translation and rotation, and of the ratio of detected outliers. The metrics of translation and rotation are defined as the mean squared forward difference of the respective translation and rotation parameter time series in their linearized Lie algebra representation. The forward difference ensures that these metrics primarily capture the subject ``activity'' during the scan: sudden bursts of motion are deemed more detrimental than slow subject drift. The outlier ratio is calculated as one minus the mean slice weight across the dataset.

To combine these three resulting metrics into one aggregate motion measure, each metric is first normalized to its respective median across the scan population. This normalization ensures equal and dimensionless weight for all metrics. The summary measure is subsequently calculated as the root-mean-squared value across the rescaled rotation, translation, and outlier metrics. This combined measure provides a relative ranking of the amount of motion across all scans, thus enabling to select scans with limited or severe motion based on their percentile rank in the cohort. For the evaluation in this paper, we selected subjects at the 5\%, 25\%, 50\%, 75\%, and 95\% quantiles of the group scanned at term (Scans~1--5).

\subsection{Validation}

In order to validate the accuracy of the proposed SHARD reconstruction, the effects of subject motion and slice dropouts are simulated in selected neonates with little motion. These subjects were selected as the least motion-affected subjects at age 31, 34, 37, 40, and 43 weeks PMA (Scans~A--E), as detected by the summary motion metric. The motion-corrected reconstructions of these scans constitute the ground truth for validating the reconstruction. In each of these, we simulate motion using the estimated motion traces of Scans~1--5, i.e., the selected percentiles of the group scanned at term. As such, we generate 25 artificial motion-corrupted datasets that span both the range of observed motion and age range in the cohort.

Simulating motion is achieved using the forward model described in section \ref{sec:forward}, including gradient reorientation and slice profile convolution. The effect of slice dropouts and spin history is simulated using multiplicative intensity scaling with a smooth bias field in every slice. These scale fields are derived as the ratio between data and prediction in Scans~1--5, smoothed in-plane to remove any residual anatomical features. Finally, we added Gaussian noise at 20\% of the noise level of the raw data, achieving data of full-rank with texture similar to the denoised images. With these steps, we obtain simulated scattered slice data with ground truth, comparable to the actual preprocessed input data. Susceptibility-induced distortion was not simulated because the B0 field is not estimated in the presented framework; including it unnecessarily complicates the validation.

SHARD reconstruction of these simulated data is run with identical parameter setup. Accuracy of the recovered motion traces is measured as the RMSE between the recovered and ground-truth motion parameters in their Lie algebra parameterization, after subtracting the global mean transformation to take out any rigid offset between independent reconstructions. Accuracy of the reconstructed data is measured as the RMSE between the output reconstruction and the ground truth projected onto the diffusion encoding, again after global rigid alignment and dMRI reorientation. The relative reconstruction accuracy is normalized to the mean $b=0$ signal in the brain mask, yielding a measure in 1/SNR units.

%%%%%%%%%%%%%%%%%%%%%%%%%%%%%%%%%%%%%%%%%%%%%%%%%%%%%%%

\section{Results}

\subsection{Validation}

The validation in simulated motion-corrupted data assesses our method's ability to solve the inverse problem. The measured accuracy of the motion parameters is listed in Tables~\ref{tbl:valmotion}a (translation) and~\ref{tbl:valmotion}b (rotation). The registration achieves an accuracy of $<\unit[0.5]{mm}$ and $<\unit[0.7]{deg}$ in the most motion-corrupted simulation, indicating the registration's capability to accurately detect and correct even severe motion. The RMSE between the recovered and simulated motion traces increases with more disruptive levels of simulated motion. This increase is linearly proportional with the measure of motion of the simulated traces. The registration accuracy is generally stable across gestational age.

\begin{table*}[t]
	\centering
	\small
	\renewcommand{\arraystretch}{1.3}
	\begin{tabular}{@{}lccccc@{}}
	\toprule
	& Scan A & Scan B & Scan C & Scan D & Scan E  \\
	\midrule
	Simulation 1 & \unit[0.10]{mm} ; $0.12^\circ$ & \unit[0.10]{mm} ; $0.13^\circ$ & \unit[0.09]{mm} ; $0.11^\circ$ & \unit[0.09]{mm} ; $0.10^\circ$ & \unit[0.09]{mm} ; $0.10^\circ$ \\
	Simulation 2 & \unit[0.12]{mm} ; $0.22^\circ$ & \unit[0.12]{mm} ; $0.21^\circ$ & \unit[0.12]{mm} ; $0.20^\circ$ & \unit[0.13]{mm} ; $0.20^\circ$ & \unit[0.12]{mm} ; $0.22^\circ$ \\
	Simulation 3 & \unit[0.17]{mm} ; $0.18^\circ$ & \unit[0.17]{mm} ; $0.17^\circ$ & \unit[0.16]{mm} ; $0.15^\circ$ & \unit[0.16]{mm} ; $0.15^\circ$ & \unit[0.18]{mm} ; $0.19^\circ$ \\
	Simulation 4 & \unit[0.21]{mm} ; $0.44^\circ$ & \unit[0.24]{mm} ; $0.51^\circ$ & \unit[0.19]{mm} ; $0.34^\circ$ & \unit[0.20]{mm} ; $0.40^\circ$ & \unit[0.21]{mm} ; $0.42^\circ$ \\
	Simulation 5 & \unit[0.43]{mm} ; $0.61^\circ$ & \unit[0.42]{mm} ; $0.62^\circ$ & \unit[0.42]{mm} ; $0.59^\circ$ & \unit[0.45]{mm} ; $0.64^\circ$ & \unit[0.42]{mm} ; $0.59^\circ$ \\
	\bottomrule
	\end{tabular}
	\caption{Accuracy of the translation and rotation parameters in the validation. These are measured as the RMSE between the simulated and recovered motion traces. Simulations 1 to 5 introduce increasing amounts of motion corruption in the data. The RMSE increases proportionally with the degree of simulated motion. All measures are stable across the 5 validation cases A--E of gestational age 31--43 wPMA.}\label{tbl:valmotion}
\end{table*}

The measured accuracy of the reconstruction is listed in Table~\ref{tbl:valrecon}. The absolute and relative RMSE of the recovered motion-corrected data increases slightly with more severe levels of simulated motion. Nevertheless, the mean relative RMSE is 1.3\%, corresponding to $\text{SNR}=77$ at $b=0$ thus indicating good accuracy of the reconstruction. We observe no consistent trend with gestational age (Scans A--E).

\begin{table*}[t]
	\centering
	\small
	\renewcommand{\arraystretch}{1.3}
	\begin{tabular}{@{}lccccc@{}}
	\toprule
	& Scan A & Scan B & Scan C & Scan D & Scan E  \\
	\midrule
	Simulation 1 & 0.424 (1.18\%) & 0.378 (1.20\%) & 0.266 (0.79\%) & 0.414 (1.23\%) & 0.516 (1.49\%)  \\
	Simulation 2 & 0.396 (1.10\%) & 0.373 (1.18\%) & 0.428 (1.27\%) & 0.401 (1.20\%) & 0.408 (1.18\%)  \\
	Simulation 3 & 0.416 (1.16\%) & 0.401 (1.27\%) & 0.400 (1.19\%) & 0.431 (1.29\%) & 0.450 (1.30\%)  \\
	Simulation 4 & 0.489 (1.36\%) & 0.427 (1.35\%) & 0.430 (1.28\%) & 0.495 (1.48\%) & 0.464 (1.34\%)  \\
	Simulation 5 & 0.528 (1.47\%) & 0.450 (1.43\%) & 0.484 (1.44\%) & 0.537 (1.60\%) & 0.551 (1.59\%)  \\
	\bottomrule
	\end{tabular}
	\caption{Accuracy of the reconstructed data in the validation. This is measured as the absolute and relative RMSE of the recovered dMRI data, after projection into the space of the uncorrupted validation dataset. All measures are consistent across the amount of simulated motion and across the 5 validation cases A--E of gestational age 31--43 wPMA.}\label{tbl:valrecon}
\end{table*}

\subsection{Test-retest experiment}

\begin{figure}[t]
	\centering
	\includegraphics[width=\columnwidth]{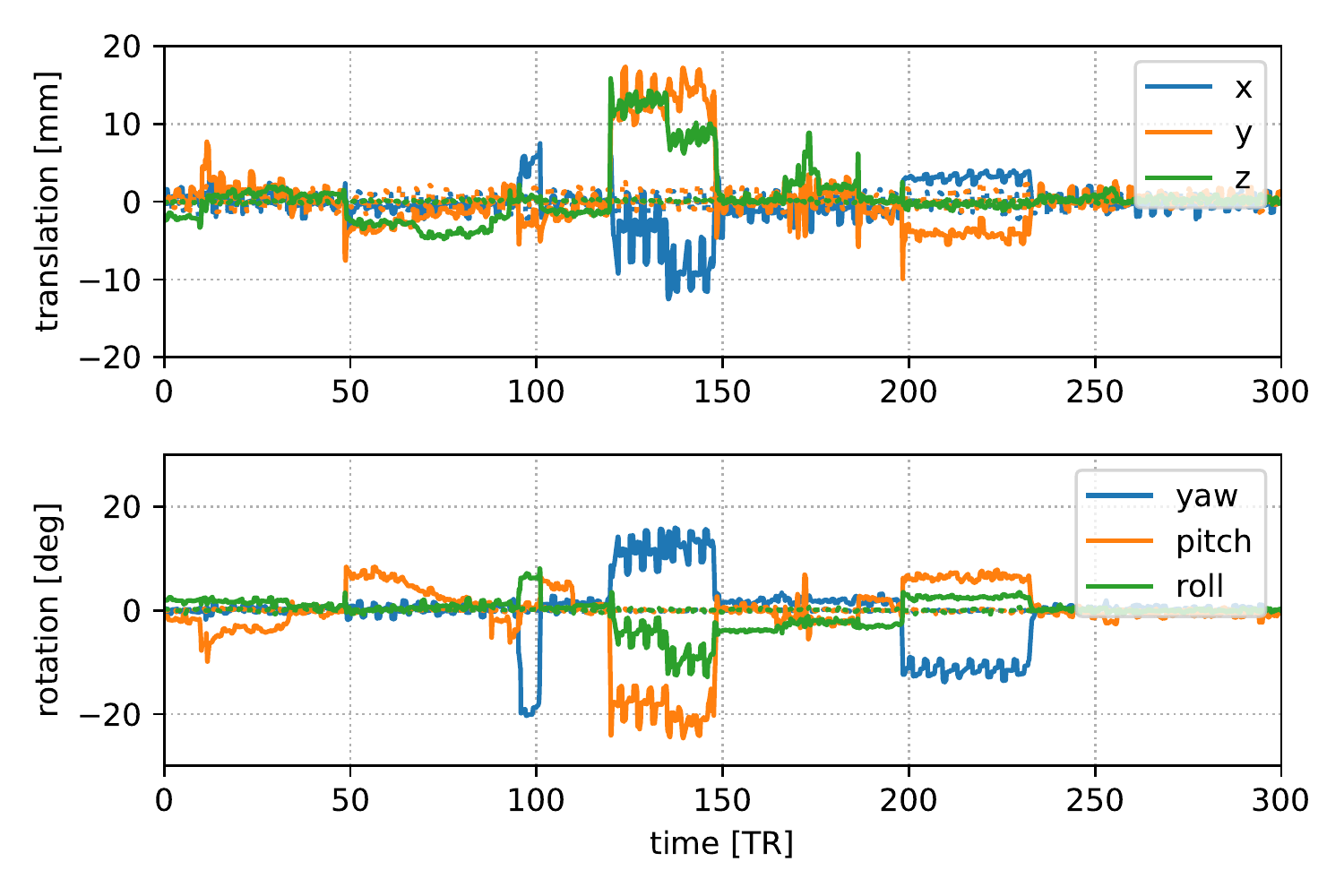}
	\caption{Estimated motion traces in the test-retest scans, plotted as dotted lines (close to zero) for the scan without deliberate subject motion and in full lines for the retest scan with motion.}\label{fig:testretest}
\end{figure}

SHARD reconstruction of the test-retest scans without and with deliberate subject motion estimated the motion traces shown in Fig.~\ref{fig:testretest}. The summary motion metric of the scan with motion is 2.71, around the 90th percentile of the motion observed in the neonatal cohort. We observed good image correspondence after symmetric registration. The mean relative RMSE is 5.4\%, corresponding to SNR = 19 at $b=0$. The median relative error is 2.6\%.

\begin{figure*}[t]
	\centering
	\includegraphics[width=\textwidth]{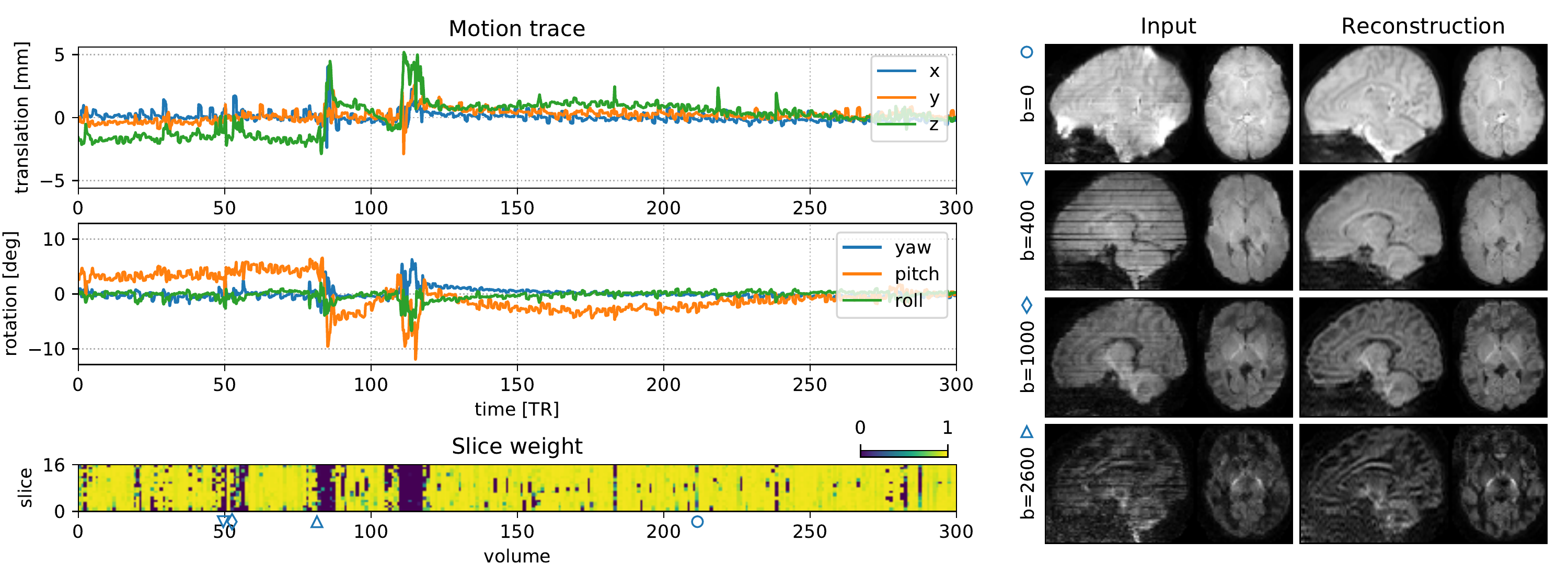}
	\caption{Motion correction output in Scan~2 (25th~percentile). The graphs on the left plot the subject motion (translation and rotation in nautical angles) over time (top), and the weight of each slice in the fit (bottom). Bursts of sudden motion coincide with increased prevalence of outlier slices. The images on the right show example volumes before and after correction. One volume per shell is shown, selected at the 10-percentile of the total slice weight per shell. The icons ($\circ$, $\triangledown$, $\lozenge$, $\vartriangle$) indicate the position of the acquired volume in the dMRI series.}\label{fig:motiontrace}
\end{figure*}

\subsection{In vivo analysis}

Figure~\ref{fig:statsplot} illustrates the distribution of the motion metrics in the population, split between preterm scans and scans at term. The histograms at the top show that the measures of subject bulk motion (translation and rotation) are heavily skewed towards the left, indicating that bulk subject motion is relatively well constrained throughout the scan. However, the ratio of slice outliers---related to short-time subject motion resulting in signal dropouts and spin history effects---is more broadly distributed between approximately 5\% and 45\%. As expected, subject translation and rotation metrics are correlated (Spearman rank correlation $\rho=0.68$ in the term group), as are the rotation and outlier metrics ($\rho=0.67$). The final histogram in the bottom right corner shows the distribution of the dimensionless summary motion metric, derived from the three metrics at the top. We observe that the detected subject motion is generally less severe in the preterm group, and more severe in scans at term. Based on the summary metric, we selected 5 scans at term, ranked at the 5\%, 25\%, 50\%, 75\%, and 95\% quantiles indicated in the figure.

The detected head motion trace of scan~2 (25th~percentile) is illustrated in Fig.~\ref{fig:motiontrace}, plotted after conversion to a conventional translation and rotation (Euler angles) description. We observe a slowly varying bulk motion, with short intermittent bursts of sudden, fast subject movement. These short bursts coincide with blocks of low slice weights, shown in the bottom panel. The outlier mixture model thus detects an increased rate of slice dropouts during periods of fast subject motion. Even for this subject with relatively limited motion (25th~percentile), we observe translation in the range of $\pm\unit[5]{mm}$ and rotation in the range of $\pm10^\circ$. In some subjects, the $x$ and $y$ translation components also contain an imprint of the alternating phase encoding in the input, due to a global offset in the field map originating from an inevitable interdependency in field map estimation \citep{Andersson2003}. Furthermore, the right panel in Fig.~\ref{fig:motiontrace} shows selected volumes (one per shell) in this scan before and after motion correction. Note how geometric distortion and intra-volume motion is corrected in the output, whilst preserving the dMRI contrast across gradient encoding direction and $b$-value.

\begin{figure*}[p]
	\centering
	\includegraphics[width=\textwidth]{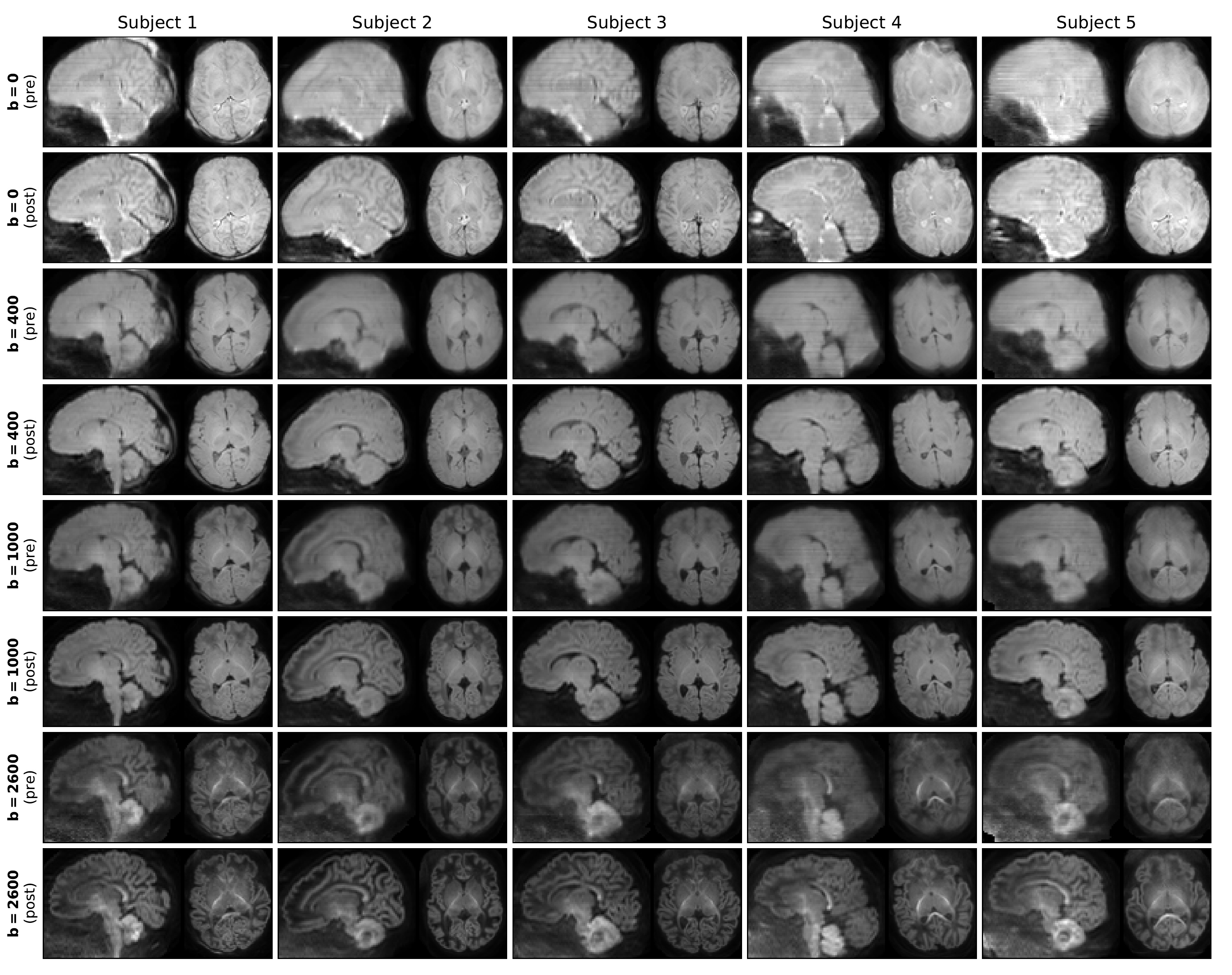}
	\caption{Mean per shell ($b$-value) in Scans 1--5, selected at the 5\%, 25\%, 50\%, 75\% and 95\% quantiles of the summary motion metric respectively. Each shell is shown before (pre) and after (post) reconstruction. We can observe a considerable gain in sharpness in the output, thanks to motion correction, field unwarping and through-slice super-resolution.}\label{fig:shellmean}
\end{figure*}

\begin{figure*}[t]
	\centering
	\begin{minipage}{0.4\textwidth}
		\centering
		%\textsf{\textbf{Scan~3 (median motion)}}\\ \vspace{1mm}
		\includegraphics[width=\textwidth]{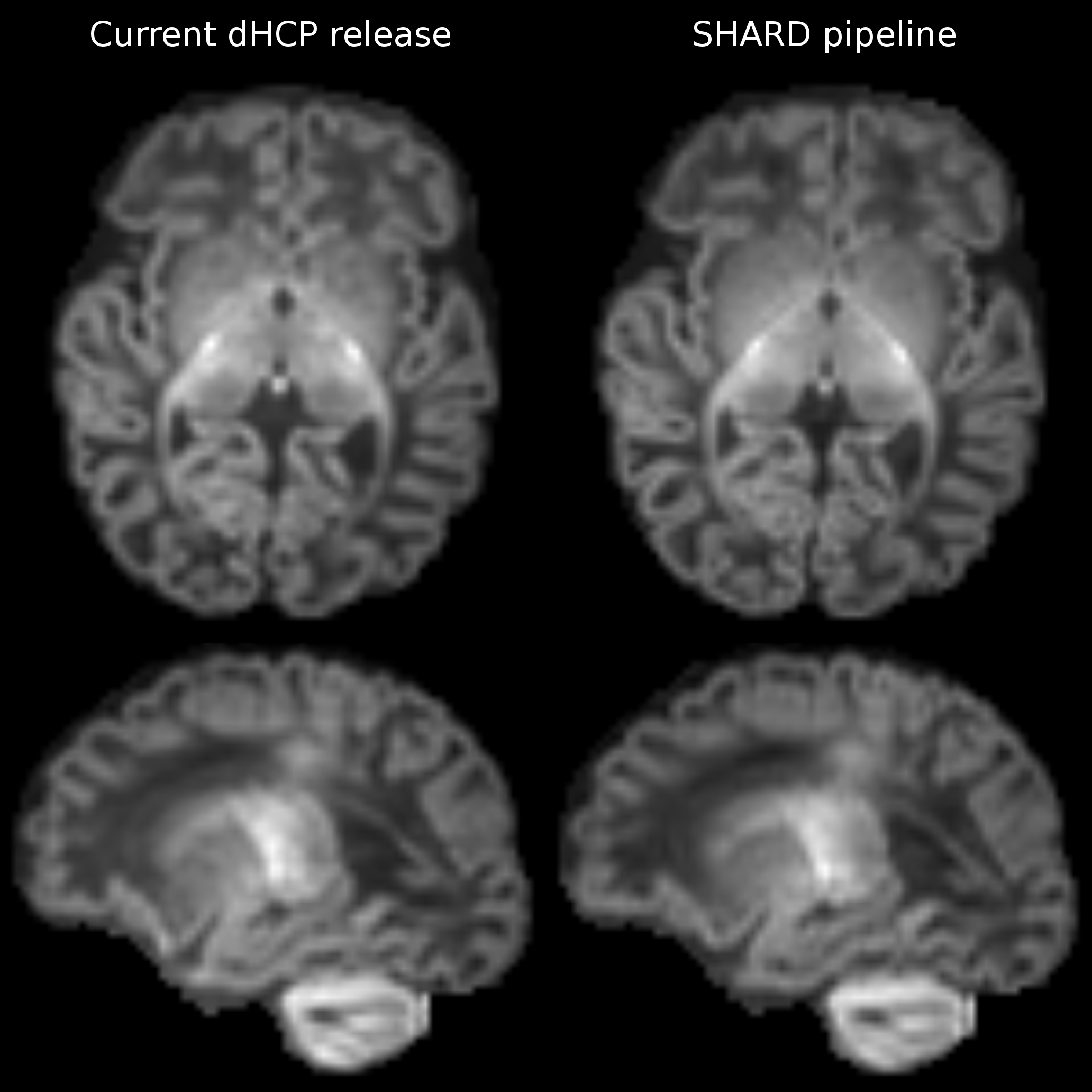}
	\end{minipage}
	%\hfill
	\begin{minipage}{0.4\textwidth}
		\centering
		%\textsf{\textbf{Scan~5 (severe motion)}}\\ \vspace{1mm}
		\includegraphics[width=\textwidth]{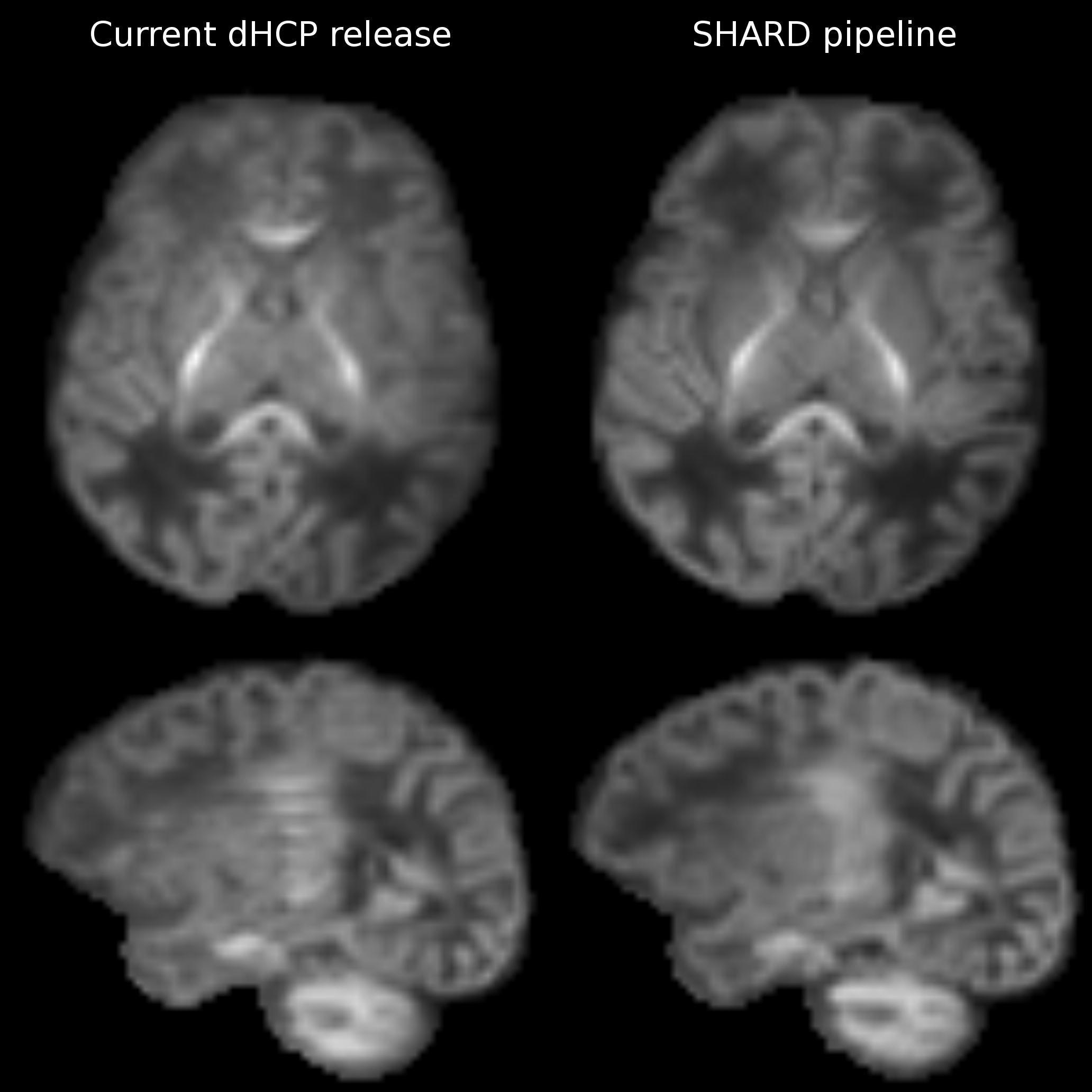}
	\end{minipage}
	\caption{Comparison of the mean $b=\unitfrac[2600]{s}{mm^2}$ shell after motion correction, when processed with the current dHCP pipeline and with SHARD-based reconstructing. The comparison is shown for Scan~3 (median motion; left) and Scan~5 (severe motion; right).}\label{fig:dhcpvsshard}
\end{figure*}

Figure~\ref{fig:shellmean} shows the spherical mean dMRI contrast per shell before and after reconstruction in all 5 selected subjects. Subject motion introduces blurring in the spherical mean, as confirmed when comparing the contrast before reconstruction between Subject~1 (limited motion) and Subject~5 (severe motion). After reconstruction, we can observe a considerable gain in image sharpness, indicating successful motion and distortion correction and effective through-slice resolution recovery. In addition, the spherical mean contrast per shell is well aligned across shells, without need for explicit inter-shell image registration. Figure~\ref{fig:dhcpvsshard} compares the mean signal in the outer shell of the SHARD pipeline to the current dHCP data release, processed with a pipeline based on FSL Eddy \citep{Andersson2017} as described in \citet{Bastiani2019}. While results are on par in Scan~3 (median motion), the SHARD reconstruction is sharper in Scan~5 (severe motion) and better corrected dropout slices. This indicates that SHARD is robust to larger ranges of subject motion than the current dHCP data release.

\begin{figure*}[t]
	\centering
	\includegraphics[width=.497\textwidth]{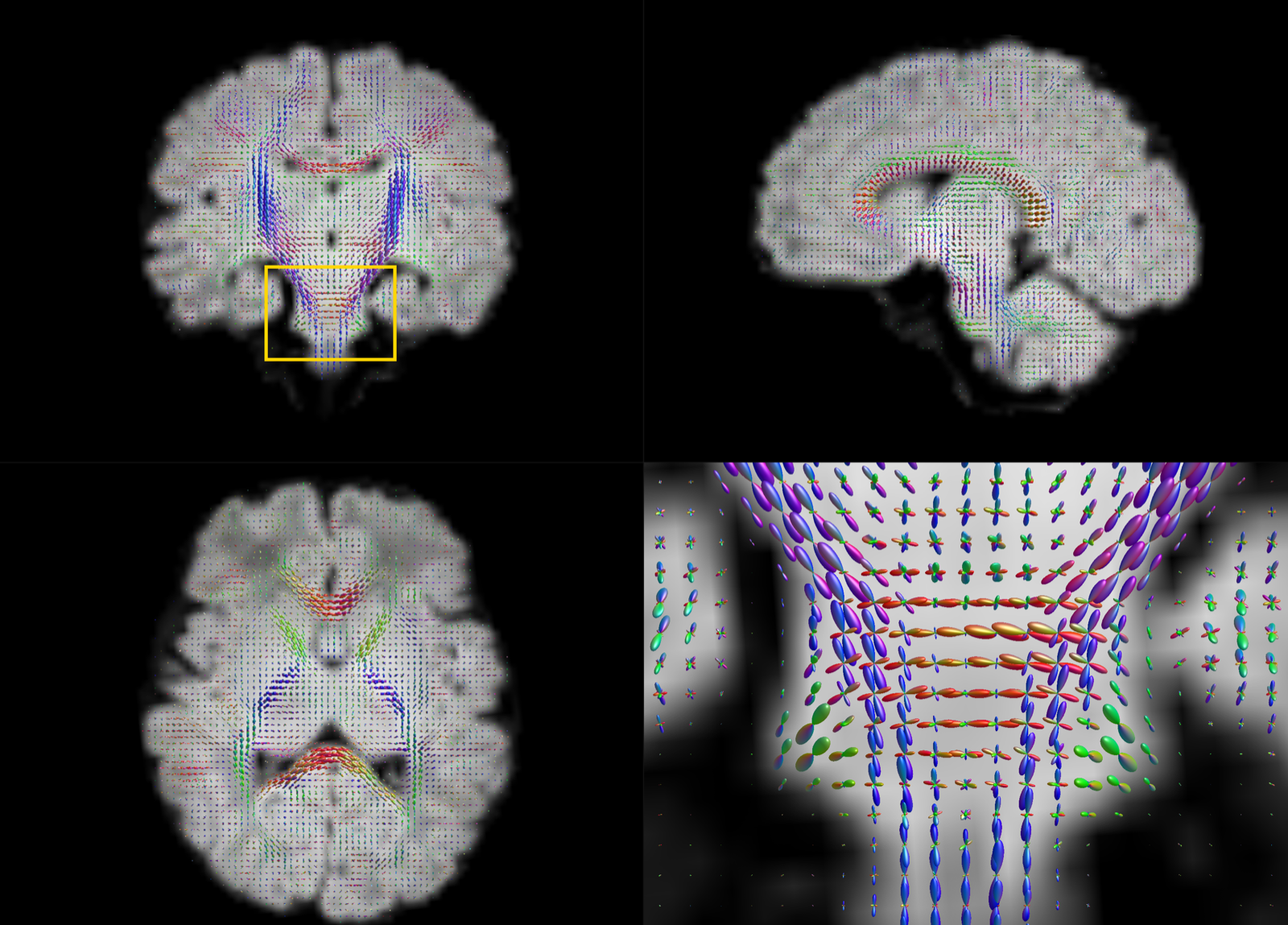}
	\hfill
	\includegraphics[width=.497\textwidth]{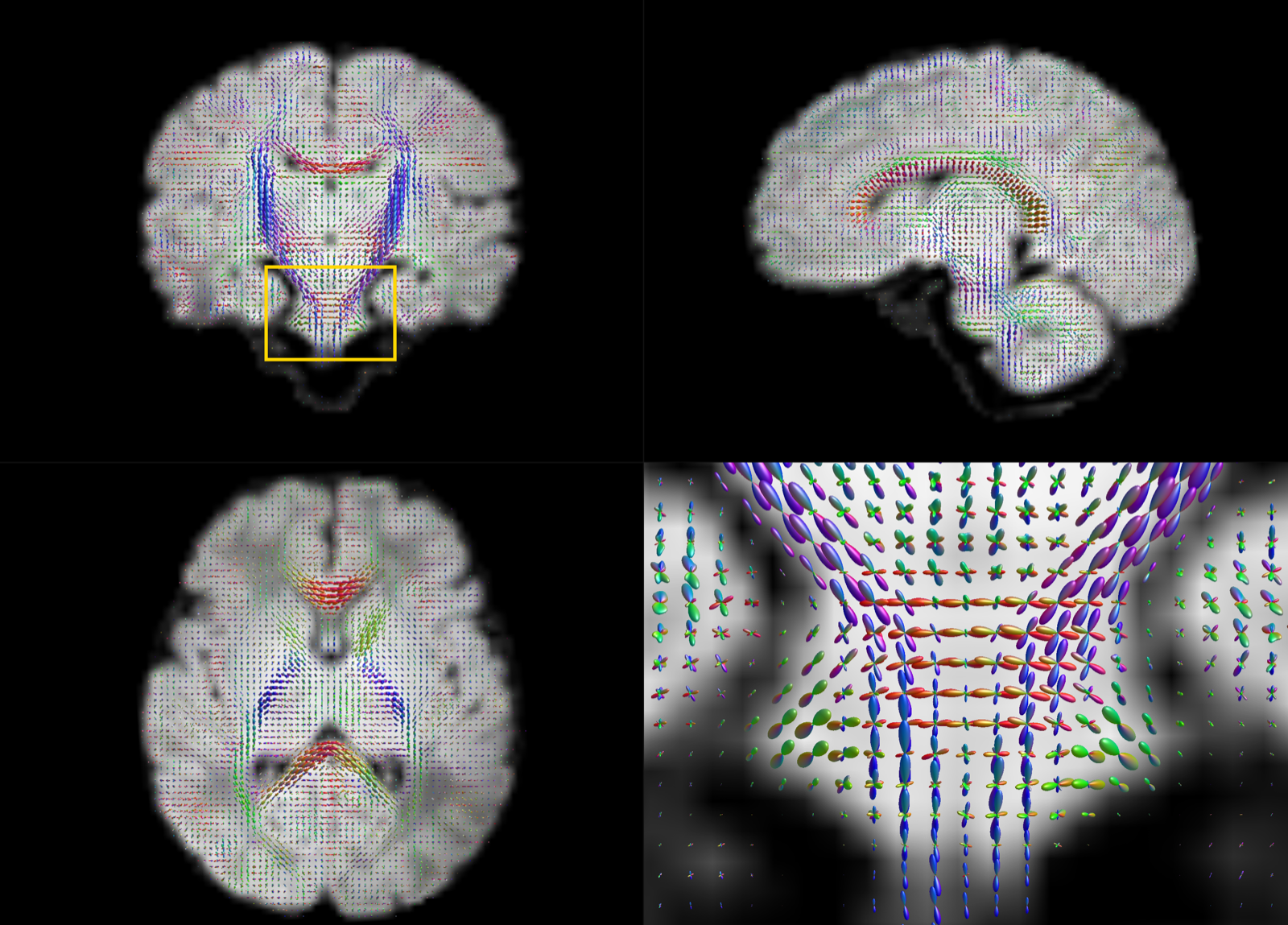}
	\caption{Estimated per-voxel tissue orientation distribution functions in Scans 1 and 5 with limited (left) and severe (right) subject motion. The enlarged region shows a coronal cross-section of the pons and the middle cerebellar peduncles.}\label{fig:odf15}
\end{figure*}

\begin{figure*}[t]
	\centering
	\includegraphics[width=.497\textwidth]{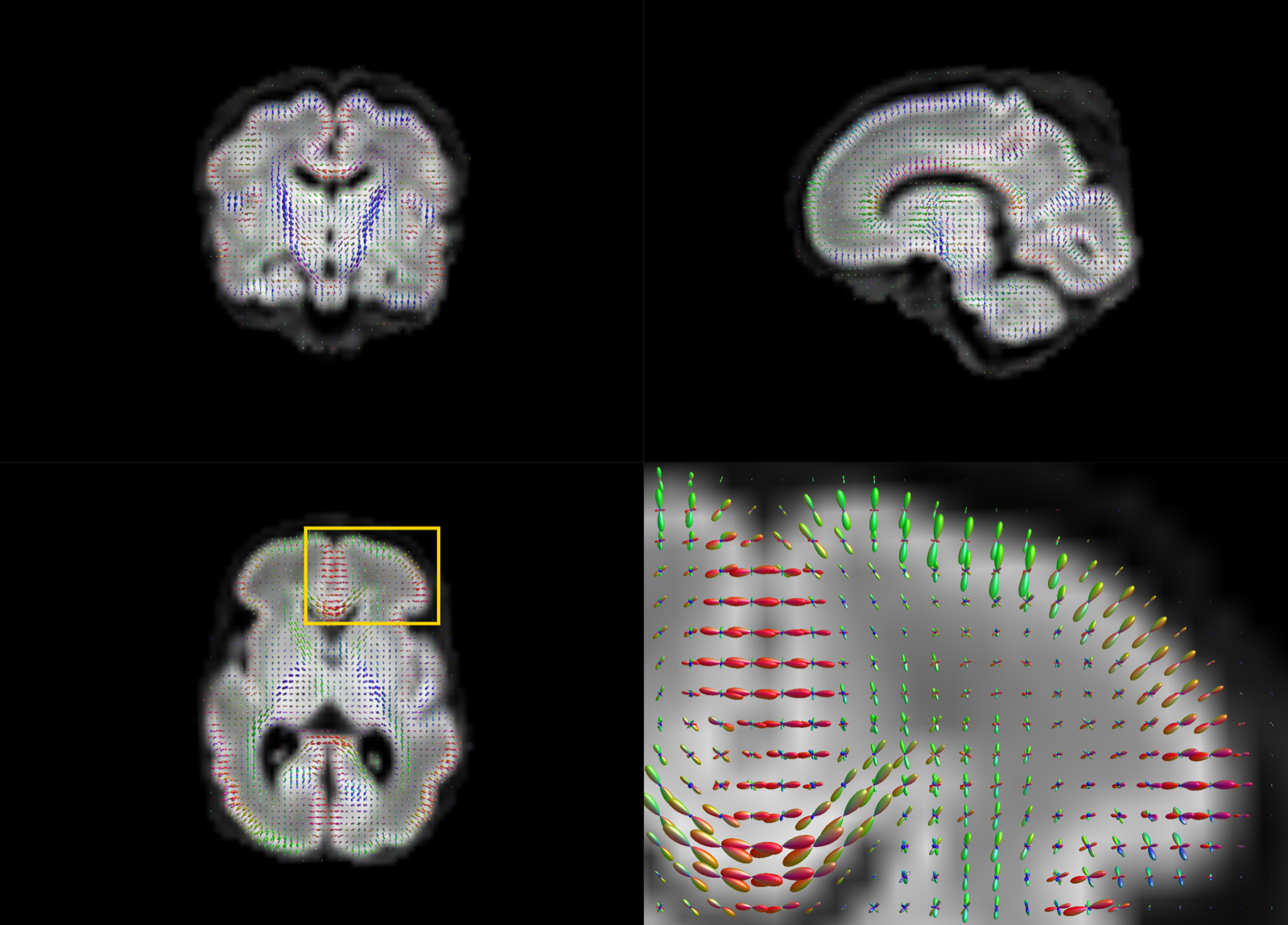}
	\hfill
	\includegraphics[width=.497\textwidth]{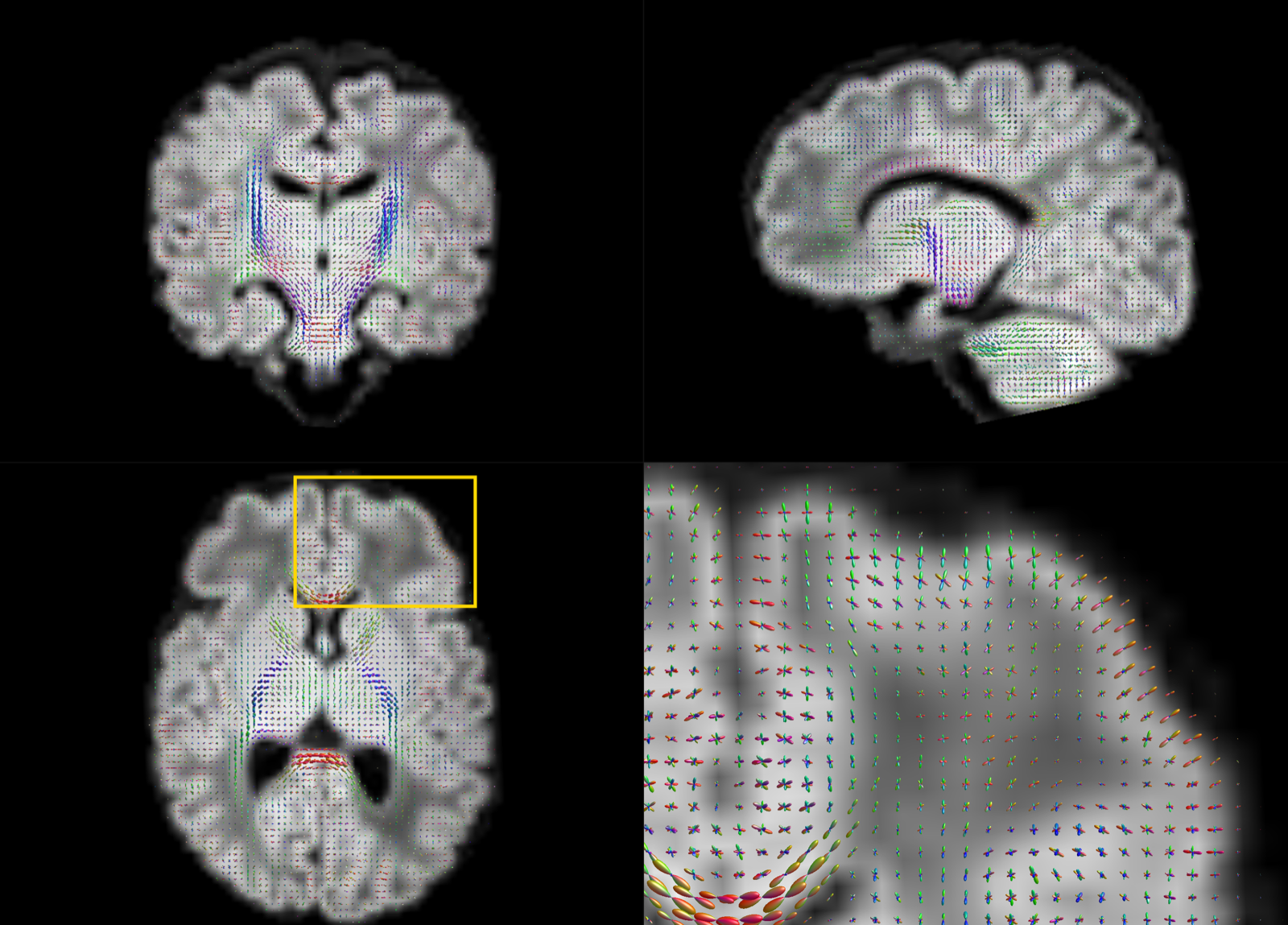}
	\caption{Estimated per-voxel tissue orientation distribution functions in preterm (\unit[31]{wPMA}; left) and at term (\unit[41]{wPMA}; right) scans of the same subject, both with close to median motion. The enlarged region shows an axial cross-section of the frontal cortex, illustrating the radial cortical structure in the early developing brain.}\label{fig:odfpreterm}
\end{figure*}

The motion-corrected super-resolved SHARD reconstructions are suitable for a wide range of analyses to explore tissue microstructure and orientation. Figure~\ref{fig:odf15} displays the tissue orientation distribution function (ODF) in each voxel, derived using a group-wise 2-component multi-shell factorization. Specifically, the 5 selected subjects at term were each decomposed into a tissue component (the SH convolution of a tissue response function and ODF at $\ell_\text{max}=8$) and an isotropic ($\ell_\text{max}=0$) fluid component, using unsupervised convexity- and nonnegativity-constrained spherical factorization \citep{Christiaens2017}. Subsequently, the 2-component spherical deconvolution \citep{Jeurissen2014} was repeated with the group-average response functions. The figure shows a comparison between scans at both ends of the motion spectrum. In both cases, we retrieve the major developing white matter tracts, including the corticospinal tracts and middle cerebellar peduncles shown in the figure. Figure~\ref{fig:odfpreterm} displays the tissue ODF in a preterm subject, scanned at birth (31~wPMA) and at term-equivalent age (41~wPMA). In this case, we can observe the radial tissue orientation in the developing cortex.

%%%%%%%%%%%%%%%%%%%%%%%%%%%%%%%%%%%%%%%%%%%%%%%%%%%%%%%

\section{Discussion}

In the presence of slice-level subject motion, dMRI volumes can contain under-sampled regions and slice dropouts, and no longer share a common diffusion encoding direction and associated diffusion-weighted contrast. These issues introduce a need for a multi-dimensional $q$-space representation, to be fitted directly to the acquired set of scattered slices (spanning all diffusion encoding $b$-values and directions). This work builds on a data-driven SHARD representation that spans both the angular and the radial $q$-space domain, using respectively the spherical harmonics basis and a bespoke singular value decomposition. Its optimal rank reduction properties are well suited for robust and multi-scale reconstruction and registration \citep{Christiaens2018}. Furthermore, its model-free nature enables fair comparison between various microstructure models and data analysis methods. The resulting motion correction framework hence reconstructs an uncorrupted SHARD representation from the scattered slice data, accounting for the varying and motion-dependent dMRI contrast and also for the slice profile to recover through-plane resolution. The estimated SHARD coefficients can be used to regenerate data with arbitrary sampling, as expected by other models, or used directly in techniques that operate on multi-shell spherical harmonics \citep{Christiaens2017, Reisert2017, Novikov2018a}. As such, the reconstructed output data can be passed on to subsequent processing pipelines for subject-level and group-level analysis of tissue miscrostructure and connectivity in the developing brain \citep{Pietsch2019, Kelly2019}.

The scattered slice perspective in which this work is rooted is a natural framework for multi-dimensional EPI reconstruction in the presence of motion and missing data. Akin to the slice shuffling techniques used in novel acquisition protocols \citep{Wu2017, Hutter2018a, Hutter2018b}, it recognises the need to break away from a concept of volume-level diffusion encoding in what is at heart a planar imaging technique. However, reconstruction of irregularly-spaced and outlier-corrupted data inevitably introduces a need for regularization, in this case applied both spatially and temporally in the reconstruction and motion estimation steps respectively. In the spatial domain, we use an isotropic Laplacian regularization term to stabilize under-sampled areas and missing data. Additionally, the reconstruction penalizes the highest frequencies in the through-plane direction otherwise amplified by the slice profile deconvolution. This term may not be so essential in isotropically sampled data, but enables us to downweight the Laplacian regularizer and thus preserve resolution in data with overlapping slices. Both regularizer weights were empirically tuned to be minimally intrusive in the reconstruction. In the temporal domain, we use a two-stage filter on subject motion trace, wherein the first stage downweights unreliable registration estimates in outlier slices and the second stage encourages piecewise temporal consistency. The latter is achieved with a median smoothing filter that preserves the sudden jumps often observed in subject motion trajectories.

Results show that the proposed slice-to-volume SHARD reconstruction accurately recovers slice-level motion and successfully corrects slice dropouts in a large cohort of neonatal data, both preterm and at term. Such slice-level motion correction is vital for reliably estimating fibre orientations and scalar microstructure parameters in neonatal and other highly motion-corrupted data. Furthermore, owing to the integrated slice profile the proposed correction also achieves through-plane super-resolution, necessary to recover the $z$-axis blurring stemming from the anisotropic acquisition. We have not observed an upper limit on the amount of resolvable motion in our neonatal cohort, and found that the relative accuracy of the estimated motion traces does not increase substantially with the simulated level of motion. In fact, the same framework has also proved effective in fetal data with more severe motion \citep{Christiaens2019ismrm, Christiaens2019}. The comparison to the current dHCP release \citep{Bastiani2019} in Figure~\ref{fig:dhcpvsshard} illustrated the robustness of SHARD reconstruction in highly motion-corrupted data. The output of the SHARD pipeline will be released as part of the next dHCP data release.

A potential risk, arising from having only a single slice orientation, is that this motion-correction framework might rely on the cross-slice link inherent to multi-band data. While we have tested the method on single band data with success, there is a risk that slice-to-volume reconstruction of a single band slice stack can lead to affine transformations (scale and shear in $z$) of the reconstruction, thus not guaranteeing anatomical accuracy. Although this risk is mitigated by the temporal motion filter, aligning the output reconstruction to an anatomical image may still be recommended for single band data. Alternatively, the reconstruction framework could be extended to support multi-view acquisitions with orthogonal slice stacks.

One limitation of the presented reconstruction is that outliers are considered shot-by-shot, whereas dMRI data can also be affected by localized sources of corruption. The slice-level outlier detection is designed to suppress slices with complete or partial signal dropout, typically due to bulk motion during the diffusion preparation \citep{Anderson1994}. However, spin history effects and physiological motion can affect the data more locally. In addition, fat shift artefacts and leakage in accelerated imaging can also introduce local outliers. An extension for voxel-based or patch-based outlier detection may alleviate the adverse effect of such localized artefacts.

Another limitation of the proposed SHARD reconstruction is that it deals only with subject motion and susceptibility-induced distortion using an invariant B0 field map. While the method is straightforward to extend to eddy current distortion by incorporating constrained affine registration, this is not currently implemented as it is not a concern in our neonatal data. Dealing with susceptibility distortion beyond the rigidly-aligned static field map is more challenging \citep{Andersson2018, Hutter2018a}. The assumption that the B0 field is invariant in the subject reference frame only holds for limited motion, and can lead to localized unresolved distortion in some subjects, typically in cases of severe motion-induced misalignment.

Future work will explore closer integration with distortion correction, extending the current rigid registration to non-linear registration along the phase encoding direction \citep{Andersson2003, Holland2010}. Moreover, such integrated distortion correction method could leverage the interleaved phase encoding scheme for dynamic susceptibility field changes \citep{Bhushan2014}. In addition, future work can expand the SHARD reconstruction framework for generalized diffusion waveforms and combined diffusion-relaxometry studies \citep{Wu2017, Hutter2018b} to enable motion correction in other multi-dimensional data with directionally-dependent contrast.

%%%%%%%%%%%%%%%%%%%%%%%%%%%%%%%%%%%%%%%%%%%%%%%%%%%%%%%

\section{Conclusion}

This work has introduced a slice-to-volume reconstruction framework for diffusion MRI, building on a data-driven SHARD representation in $q$-space that aims to align the acquired scattered slices both within and across shells. The framework is applied in neonatal dHCP data, demonstrating successful motion correction at different motion levels and ages at scan as present in the cohort, but can be equally useful in fetal, pediatric and adult dMRI.

%%%%%%%%%%%%%%%%%%%%%%%%%%%%%%%%%%%%%%%%%%%%%%%%%%%%%%%

\section*{Acknowledgements}
\begin{footnotesize}
This work received funding from the European Research Council under the European Union's Seventh Framework Programme [FP7/20072013], ERC grant agreement no. [319456] (developing Human Connectome Project) and MRC strategic funds [MR/K006355/1]. D.C. is supported by the Flemish Research Foundation (FWO), fellowship no. [12ZV420N]. J.H. is supported by Wellcome Trust Fellowship [WT/201374/Z/16/Z]. This work was also supported by the Wellcome/EPSRC Centre for Medical Engineering at King's College London [WT~203148/Z/16/Z] and by the National Institute for Health Research (NIHR) Biomedical Research Centre at Guy's and St Thomas' NHS Foundation Trust and King's College London. The views expressed are those of the authors and not necessarily those of the NHS, the NIHR or the Department of Health.\par
\end{footnotesize}

%%%%%%%%%%%%%%%%%%%%%%%%%%%%%%%%%%%%%%%%%%%%%%%%%%%%%%%

\section*{References}

\bibliographystyle{model2-names}		% with doi
\bibliography{references}

%%%%%%%%%%%%%%%%%%%%%%%%%%%%%%%%%%%%%%%%%%%%%%%%%%%%%%%
% SUPPLEMENTARY MATERIAL

\end{document}